\documentclass[cits]{PoS}

\usepackage{amsmath}
\usepackage{amsfonts}
\usepackage{graphicx}

\title{Longitudinal thermalization via the chromo-Weibel instability}

\ShortTitle{Longitudinal thermalization via the chromo-Weibel instability}

\author{\speaker{M. Attems},$^{ab}$ A. Rebhan$^a$, and M. Strickland$^{bc}$\\
\llap{$^a$}Institut f\"ur Theoretische Physik, Technische Universit\"at Wien\\
                Wiedner Hauptstrasse 8-10, 1040 Vienna, Austria.\\
\llap{$^b$}Frankfurt Institute for Advanced Studies\\
		Ruth-Moufang-Str. 1, 60438 Frankfurt am Main, Germany.\\
\llap{$^c$}Department of Physics, Kent State University\\
		Kent, OH 44242, USA.\\
}
\abstract{Non-Abelian plasma instabilities play an important role in the non-equilibrium dynamics of a weakly coupled quark-gluon plasma. Using the discretized hard loop framework we calculate the time evolution of soft gluonic fields in a longitudinally free streaming background. Extrapolating our results to energies probed in relativistic heavy-ion collisions we find a pressure anisotropy that persists for a few fm/c. However, the chromofields quickly develop a Boltzmann longitudinal energy spectrum, suggesting fast longitudinal thermalization of the quark gluon plasma even though it remains momentum-space anisotropic.  In this proceedings contribution we review our recent numerical results, present new results for the scaling of the isotropization time with the initial current fluctuation amplitude, and present tests of the gauge invariance of the extracted longitudinal spectra.}
\FullConference{Xth Quark Confinement and the Hadron Spectrum,\\
		October 8-12, 2012\\
		Technische Universit\"at M\"unchen, Campus Garching, Munich, Germany}
\date{\today}

\begin{document}

\section{Introduction}

The determination of the thermalization and isotropization time a quark gluon plasma (QGP) produced in relativistic heavy ion collisions is an important open question.  In this proceedings contribution we review recent numerical simulations which use the hard-expanding-loop framework for describing the evolution of non-Abelian plasma instabilities in a free-streaming background \cite{Attems:2012js} where we observed the appearance of a Boltzmann distribution of the energy in longitudinal wave number. In addition, we present two extensions of previous results:  (a) results on the scaling of the plasma isotropization time with the initial current fluctuation amplitude and (b) test the gauge invariance of the extracted longitudinal spectra subject to random local gauge transformations.  We find that the isotropization time scales with $(\log\Delta^{-1})^2$ with $\Delta$ being the initial current fluctuation amplitude.
The quasi-thermal longitudinal energy spectra, while being inherently gauge dependent, turn out to be reasonably robust against random gauge transformations as long as the latter do not introduce uncomfortably large numerical errors in the gauge invariant sum over the energy modes.
\section{The Hard Expanding Loop Framework}

The investigation of the evolution of soft (gauge) fields subject to dynamical instabilities such as the chromo-Weibel instability \cite{Heinz:1985vf,Mrowczynski:1988dz,Pokrovsky:1988bm,Mrowczynski:1993qm,Blaizot:2001nr,Romatschke:2003ms,Arnold:2003rq,Arnold:2004ih,Romatschke:2004jh,Arnold:2004ti,Mrowczynski:2004kv,Rebhan:2004ur,Rebhan:2005re} is an active area of research.  Field dynamics in an expanding background have been recently investigated using classical Yang-Mills simulations \cite{Romatschke:2005pm,Romatschke:2006nk,Fukushima:2011nq,Berges:2012iw}, analytically in the high energy limit \cite{Kurkela:2011ti,Kurkela:2011ub}, within scalar $\phi^4$ theory subject to parametric resonance instabilities \cite{Dusling:2012ig}, and SU(2) Vlasov-Yang-Mills \cite{Romatschke:2006wg,Rebhan:2008uj,Attems:2012js} including longitudinal expansion.  In this contribution we will report on the last category of investigations, namely those performed using SU(2) Vlasov-Yang-Mills simulations in a longitudinally free-streaming background.  The simulations using this framework rely on the hard-loop approximation which is valid at asymptotically weak coupling  and provide information on the evolution of the quark gluon plasma in the high energy limit.  The resulting analytical and numerical framework has been dubbed the hard expanding loop (HEL) framework.  Below we briefly summarize the key points relevant for such studies and refer the reader to Ref.~\cite{Attems:2012js} for more details.

In this framework we assume that the background particles are longitudinally free streaming and as a result the background (hard) particles possess a local rest frame momentum-space anisotropy which increases monotonically in proper-time as $\xi(\tau) = (\tau/\tau_{\rm iso})^2 - 1$, where $\tau_{\rm iso}$ corresponds to the proper time at which the distribution is originally isotropic (when $\tau_{\rm iso}<\tau_0$ with $\tau_0\sim Q_s^{-1}$, the hard particle distribution starts already anisotropic). With any isotropic distribution $f_{\rm iso}$, the free streaming background distribution, $f_0$, is constructed as
\begin{equation}
\label{faniso}
f_0(\mathbf p,x)=f_{\rm iso}\!\left(\!\sqrt{p_\perp^2+(\frac{p'^z\tau}{\tau_{\rm iso}})^2}\!\right)\!
=f_{\rm iso}\!\left(\!\sqrt{p_\perp^2+ p_\eta^2/\tau_{\rm iso}^2}\!\right) .
\end{equation}
Following \cite{Attems:2012js} we begin with the Vlasov-type dynamical equation which is obeyed by color perturbations $\delta\!f^a$ of a color-neutral longitudinally free-streaming expanding momenta distribution $f_0$
\begin{align}
V\cdot  D\, \delta\!f^a\big|_{p^\mu}=g  V^\mu
F_{\mu\nu}^a  \partial_{(p)}^\nu f_0(\mathbf
p_\perp, p_\eta) \, .
\label{Vlasov}
\end{align}
Eq.~(\ref{Vlasov}) has to be solved simultaneously with the non-Abelian Yang-Mills equations which couple the color-charge fluctuations back to the gauge fields via the induced color-currents $j^\nu_a$
\begin{align}
D_\mu  F^{\mu \nu}_a  = j^\nu_a = 
g\, t_R \int{\frac{d^3p}{ (2\pi)^3}} \frac{p^\mu}{2 p^0} \delta\!f_a(\mathbf p,\mathbf x,t) \,. 
\label{Maxwell}
\end{align}
The above equations are then transformed to comoving coordinates with the metric $ds^2=d\tau^2-d\mathbf x_\perp^2-\tau^2 d\eta^2=g_{\alpha\beta}dx^\alpha dx^\beta$.  In these coordinates the four-velocity $V^\alpha$ of hard particles has the form 
\begin{align}
 V^\alpha = \frac{ p^\alpha}{p_\perp} =
\left(\cosh (y-\eta),\,\cos\phi,\,\sin\phi,\,\frac{1}{\tau}\sinh (y-\eta)\right),
\label{velocityDef1}
\end{align}
and, as usual, $D_\alpha=\partial_\alpha-ig[ A_\alpha,\cdot]$ is the gauge covariant derivative and $F_{\alpha\beta}=\partial_\alpha  A_\beta-\partial_\beta  A_\alpha -ig[ A_\alpha, A_\beta]$ is the field strength tensor with strong coupling $g$.  To numerically solve the resulting set of coupled non-linear partial differential equations, which are expressed compactly via Eqs.~(\ref{Vlasov}) and (\ref{Maxwell}), we use temporal gauge $A^\tau = 0$ and introduce gauge-field canonical momenta
\begin{align}
\Pi^{i} = \tau \partial_\tau A_{i}
\qquad \qquad
\Pi^\eta=\frac{1}{\tau}\partial_\tau A_\eta \, .
\end{align}
The canonical momenta are related to the transverse and longitudinal chromoelectric field components via $E_i = \Pi_i/\tau$ and $E_L = \Pi^\eta$, respectively.  One can express the equations of motion in terms of canonical momenta $\Pi$ and gauge links $U$ with $U_\mu$ being the parallel transporter in $\mu$ direction:
\begin{align}
U_i = {\rm exp}(-i A^i) 
\qquad \qquad
U_\eta = {\rm exp}(i a_\eta A_\eta) \, .
\end{align}
%
%Both $\Pi$'s and $U$'s live on links between the lattice sites.  
In addition, one introduces auxiliary fields ${\cal W}(\tau,x;\phi,y)$ which describe the induced current fluctuations at each point $x$ of the three-dimensional space at given proper time $\tau$.  At each space-time point these auxiliary fields also depend on the light-like velocity of hard particles specified by the azimuthal angle $\phi$ and momentum-space rapidity $y$.  For details we refer the reader to Ref.~\cite{Attems:2012js}.

\section{Initial Current Fluctuations}

%%%%%%%%%%%%%%%%%%%%%%%%%%%%%%%%%%%%%%%%%%%%%%%%%%%%%%%%
\begin{figure}[t]
\hspace{4cm}
{\footnotesize (a)}
\hspace{7cm}
{\footnotesize (b)}\\
\includegraphics[width=0.5\textwidth]{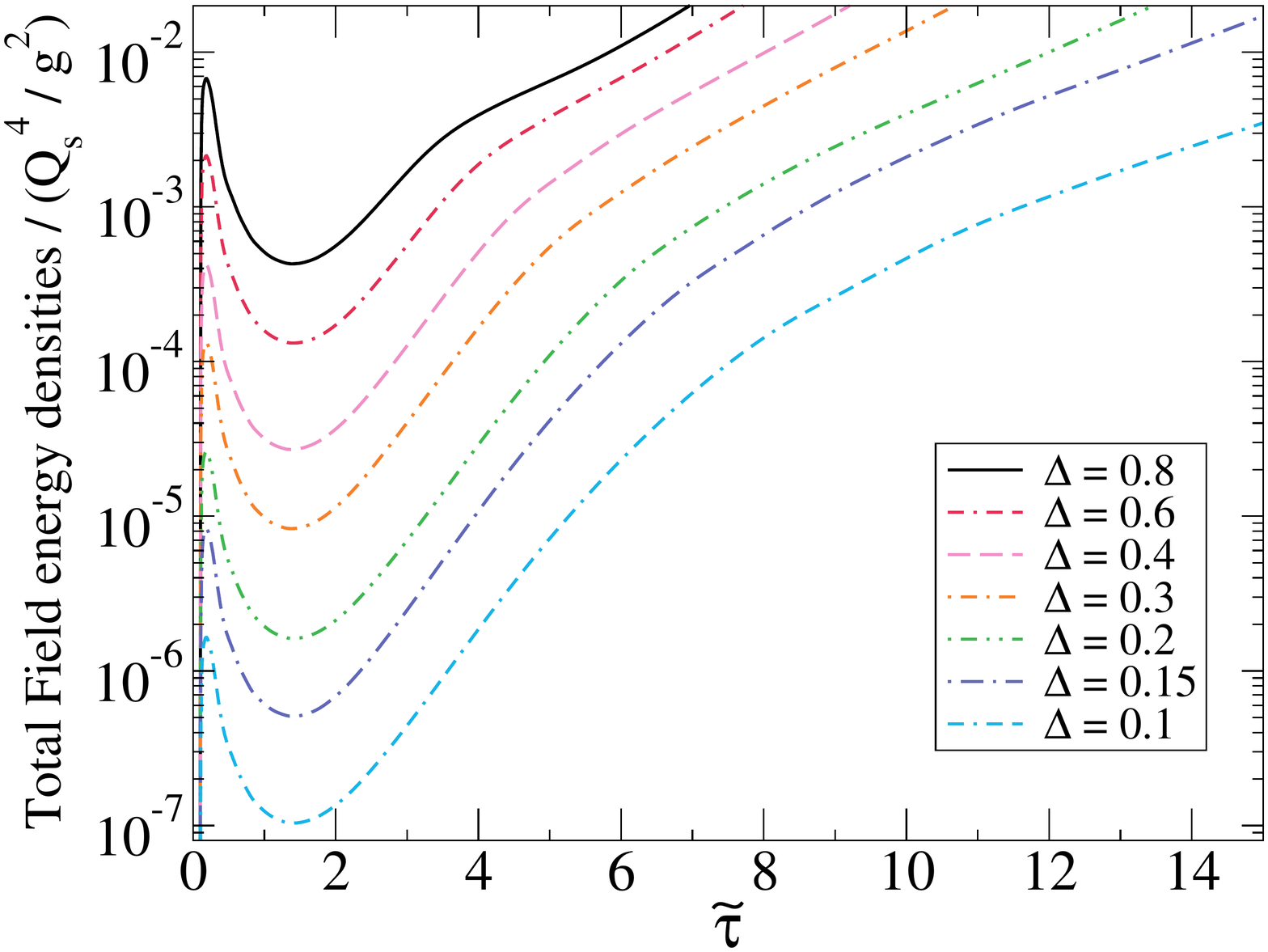}
\includegraphics[width=0.5\textwidth]{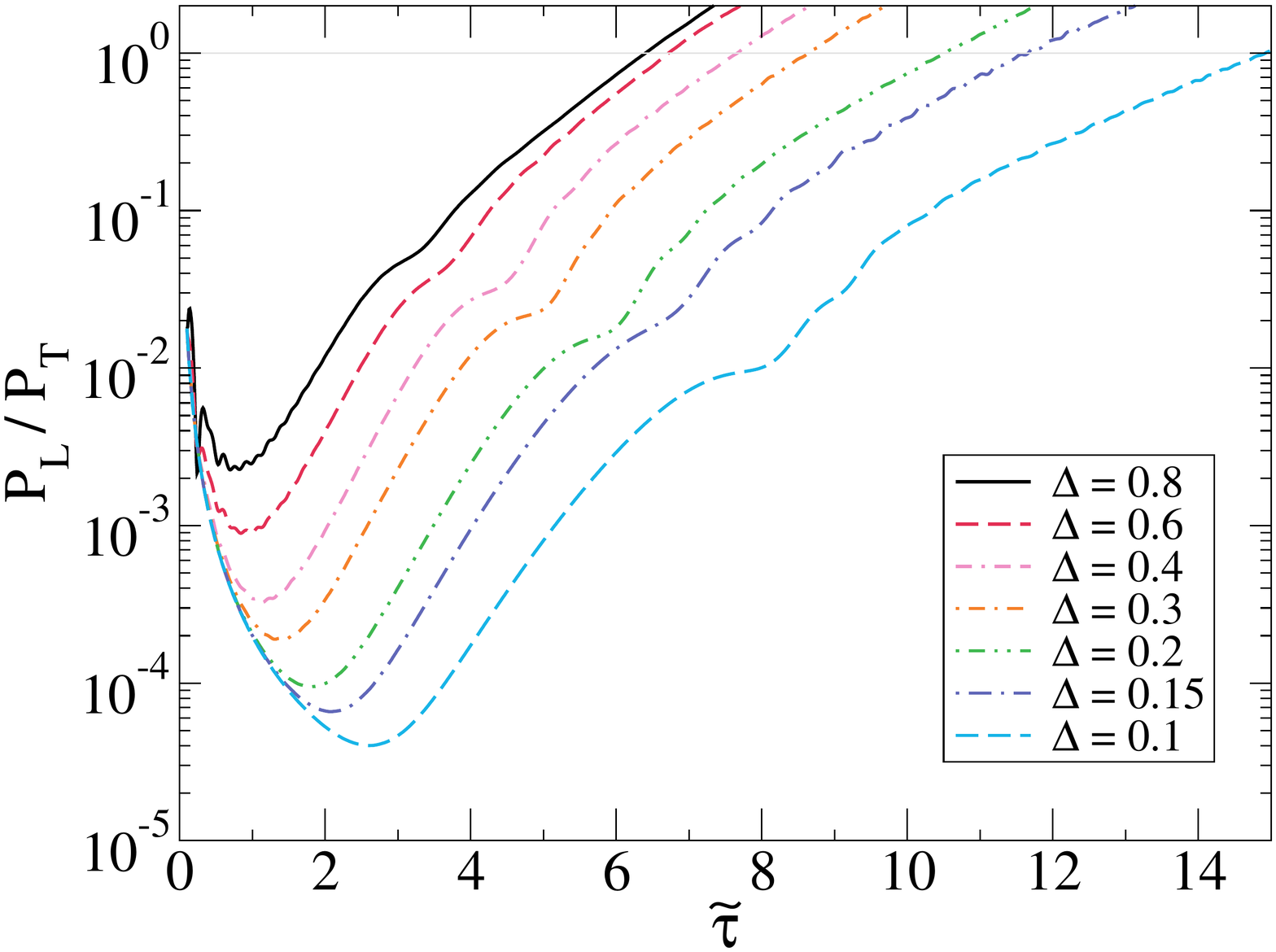}

\caption{%(Color online)
The left plot (a) shows the total field energy density for different initial
current fluctuation magnitudes $\Delta \in \{0.1,0.15,0.2,0.3,0.4,0.6,0.8\}$. The right plot
(b) features the total longitudinal pressure over the total transverse pressure
as a function of proper time again for different initial current fluctuation
magnitudes $\Delta \in \{0.1,0.15,0.2,0.3,0.4,0.6,0.8\}$. For both plots the data were taken
from the same runs.
}
\label{fig:eDensitypressure}
\end{figure}
%%%%%%%%%%%%%%%%%%%%%%%%%%%%%%%%%%%%%%%%%%%%%%%%%%%%%%%%

One of the key questions which we wish to address with HEL simulations is the question of how long does it take for the system to become isotropic in momentum space when particle and field contributions are combined. %The transverse and longitudinal pressures can be read directly from energy-momentum tensor and both contain two distinct contributions coming from the particle and field degrees of freedom.  
Due to the fact that the hard particles are assumed to be undergoing longitudinal free-streaming, they will have a  longitudinal pressure which decreases asymptotically as $\tau^{-3}$ and a transverse pressure which decreases asymptotically as $\tau^{-1}$.  In response to this growing momentum space anisotropy unstable chromofields grow rapidly after some initial transient period which depends, in part, on details of the initial spectrum.  At late times, one finds that the field pressures become large and are able to completely compensate for the anisotropy present in the particle sector.  As a result, at some point in time, the total (particle + field) pressure will be become isotropic with ${\cal P}_T = {\cal P}_L$.  One can extract this time scale directly from the simulations and, as shown in Ref.~\cite{Attems:2012js}, one finds that the resulting isotropization time-scale $\tau_{\rm ISO}$ 
(not to be confused with the initial $\tau_{\rm iso}$)
primarily depends on the initial current fluctuation amplitude $\Delta$ used as a seed as specified in Ref. ~\cite{Rebhan:2009ku}.

In Fig.~\ref{fig:eDensitypressure} we present the field energy density (left) and the ratio of longitudinal to transverse pressures (right) as a function of proper time.\footnote{For LHC and RHIC initial energy densities one unit in $\tilde\tau$ corresponds to approximately 1 fm/c and 2 fm/c, respectively.}   From the right panel we see that after an initial transient period, there is a rapid growth in the chromofield energy density.  In both plots the black solid curves are averages over 50 runs whereas the various colored dashed curves are taken from averages over 10 runs.\footnote{All runs have the lattice size $N_T^2 \times N_\eta \times N_u \times N_\phi = 40^2 \times 128 \times 128 \times 32$ with transverse lattice spacing of $a = Q_s^-1$ and  longitudinal lattice spacing of
$a_\eta = 0.025$.}
From the results contained in the right panel, one can extract the scaling of the isotropization time by extracting the proper time at which ${\cal P}_L/{\cal P}_T = 1$ for the different assumed values of $\Delta$.  The data shown were obtained from new runs of our simulation code which has been optimized through further parallelization so that we can efficiently investigate the dynamics at late times. % (see Appendix A for brief description of the code improvements).  

One caveat concerning the extracted isotropization times is that the HEL framework does not fully 
incorporate the back-reaction of the hard particles on the soft chromofields 
(it does so only in the limit of infinitesimally small scattering angles $\theta \sim g$, and thus is neglected in $f_0$).  
In addition, the hard particles are treated as unlimited energy 
reservoir from which the soft fields can obtain energy.  For these reasons, the late time behavior of the system, when the energy density in the fields becomes comparable to that contained in the particles, is no longer reliable.  Hence, one should take the inferred isotropization times as a lower bound on the time where complete isotropization could take place.

\begin{figure}[t]
\centerline{\includegraphics[width=0.5\textwidth]{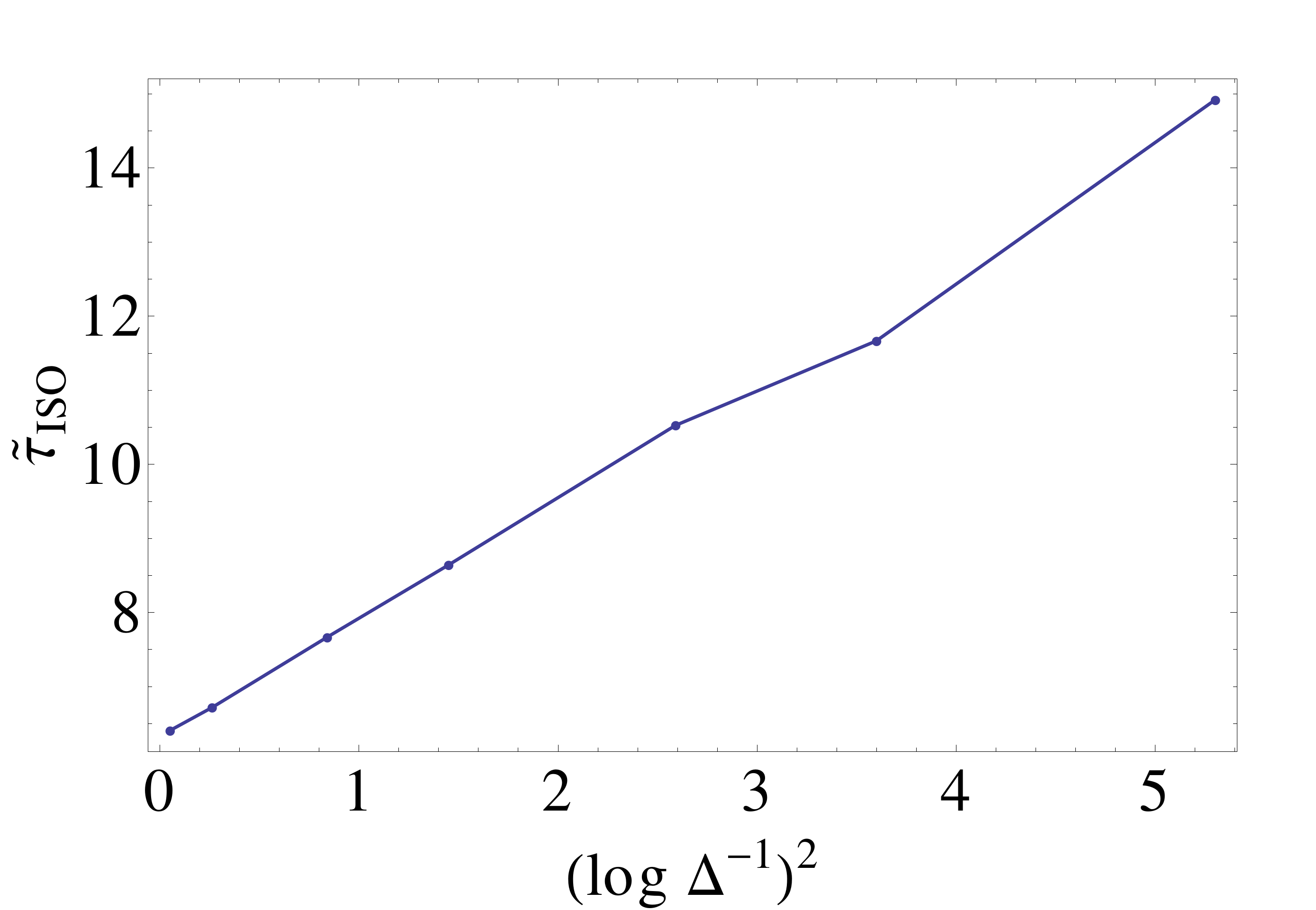}}
\caption{%(Color online)
The isotropization time $\tilde\tau_{\rm ISO}$ plotted versus $(\log\Delta^{-1})^2$.
The data is taken from the runs of Fig.~\protect\ref{fig:eDensitypressure}.
}
\label{fig:deltascaling}
\end{figure}

In Fig.~\ref{fig:deltascaling} we plot the extracted isotropization times as a function of $(\log\Delta^{-1})^2$.  As can be seen from this figure, one finds that the isotropization time increases with decreasing initial current fluctuation amplitude.  As the figure shows, we find a scaling consistent with $\tau_{\rm ISO} \sim (\log\Delta^{-1})^2$.  Our data is restricted to the set of $\Delta$'s shown in Fig.~\ref{fig:eDensitypressure}.  As explained in Ref.~\cite{Attems:2012js}, in order to generate  gauge-field occupation numbers $\sim 1/2$ consistent with those expected from initial quantum-mechanical rapidity fluctuations \cite{Fukushima:2006ax} one should choose $\Delta \sim 1.6$.  Unfortunately, due to numerical limitations stemming from the fact that we simulate compact gauge groups, we are currently unable to use such a large value of $\Delta$ and must extrapolate to the appropriate initial current fluctuation amplitude of $\Delta = 1.6$.  Based on the results shown in  Fig.~\ref{fig:eDensitypressure} and extrapolating to $\Delta = 1.6$, we obtain $\tilde\tau_{\rm ISO} \sim 5$.

%%%%%%%%%%%%%%%%%%%%%%%%%%%%%%%%%%%%%%%%%%%%%%%%%%%%%%%%
%%%%%%%%%%%%%%%%%%%%%%%%%%%%%%%%%%%%%%%%%%%%%%%%%%%%%%%%

\section{Energy Spectra}

The total longitudinal spectra are obtained following Ref.\
\cite{Fukushima:2011nq} by first Fourier transforming each field component
$E_\perp(x_\perp, \eta)$, $E_\eta(x_\perp, \eta)$, $B_\perp(x_\perp, \eta)$ and
$B_\eta(x_\perp, \eta)$, integrating over the transverse wave vectors and
decomposing each according to the longitudinal wave vector $\nu$,
in terms of which the electric and magnetic energy densities
are decomposed into longitudinal energy spectra,
\begin{align}
{\mathcal E}_E &= \int \frac{d\nu}{2 \pi} {\mathcal E}_E(\nu)  =  \int \frac{d\nu}{2 \pi} [ {\mathcal E}_{E_L}(\nu) + {\mathcal E}_{E_T}(\nu) ]\, , \\
{\mathcal E}_B &= \int \frac{d\nu}{2 \pi} {\mathcal E}_B(\nu)  =  \int \frac{d\nu}{2 \pi} [ {\mathcal E}_{B_L}(\nu) + {\mathcal E}_{B_T}(\nu) ]\, ,
\label{eq:nusum}
\end{align}
which is numerically crosschecked by comparing with the directly calculated
total energy density.

The spectral decomposition in $\nu$ thus defined
is not gauge invariant; only the integral over $\nu$ is. 
The standard procedure to extract mode spectra
is to cool down by means of Coulomb lattice gauge fixing \cite{Moore:1997cr,Ipp:2010uy}, which minimizes unphysical
high-momentum noise within 3-dimensional time slices.
However, since the left-hand-side of (\ref{eq:nusum}) is gauge
invariant, the possible redistribution by means of gauge transformations
of the fields is constrained, and we expect gauge dependencies
to be be milder than those encountered in conventional
mode spectra prior to Coulomb gauge fixing.

%By
%computing the total energy density of all modes for each time step we
%crosschecked the spectra composition. This indicates control over the
%unstable modes, which are highly populated in the infrared
%sector. Indeed the gauge invariance of the integral Eq.~(\ref{eq:nusum}) limits
%the possible redistribution due to the left gauge freedom of the used temporal
%axial gauge. Nevertheless standard procedure is to cool down the spectra thanks
%to Coulomb lattice gauge fixing \cite{Moore:1997cr,Ipp:2010uy} and extract
%the  distribution functions of the modes. This minimizes unphysical
%high-momentum noise within 3-dimensional time slices.

In order to check the robustness of the 
energy spectra against gauge issues we have performed 
random gauge transformations
$\tilde U(x)$ which produces arbitrary noise
in pure-gauge modes. The integrated energy which is manifestly gauge invariant
must stay invariant under this transformation.
However, because of lattice discretization there will be
numerical errors in the mode spectra. We have made the
amplitude of the random gauge transformations as large as possible
without violating the invariance of the total energy beyond
the percent level (with occasional deviations of up to 2\%),
which restricted the amplitude of the random SU(2) gauge parameters 
$\xi$ in $\tilde U(x)=\exp(i\vec\xi(x)\cdot\vec\tau/2)$ to $|\vec\xi|_{\rm r.m.s.}\le 0.2$.

%In the following equations of the gauge transformations we implicitly
%introduce rescaled lattice variables according to \cite{Attems:2012js}
%%
%\begin{align}
%\bar U_{\alpha,s+\frac{1}{2}} (\tau) = \tilde U^\dagger_{\alpha,s+\frac{1}{2}} (\tau)
%U_{\alpha,s+\frac{1}{2}} (\tau) \tilde U_{\alpha,s+\frac{3}{2}} (\tau)
%\end{align}
%\begin{align}
%\bar \Pi_{\alpha,s+\frac{1}{2}} (\tau) =
%\tilde U^\dagger_{\alpha,s+\frac{1}{2}} (\tau)
% \Pi_{\alpha,s+\frac{1}{2}} (\tau)
%\tilde U_{\alpha,s+\frac{3}{2}} (\tau)
% \,.
%\end{align}

%%%%%%%%%%%%%%%%%%%%%%%%%%%%%%%%%%%%%%%%%%%%%%%%%%%%%%%%
\begin{figure}[t]
\hspace{4cm}
{\footnotesize (a)}
\hspace{7cm}
{\footnotesize (b)}\\
\includegraphics[width=0.5\textwidth]{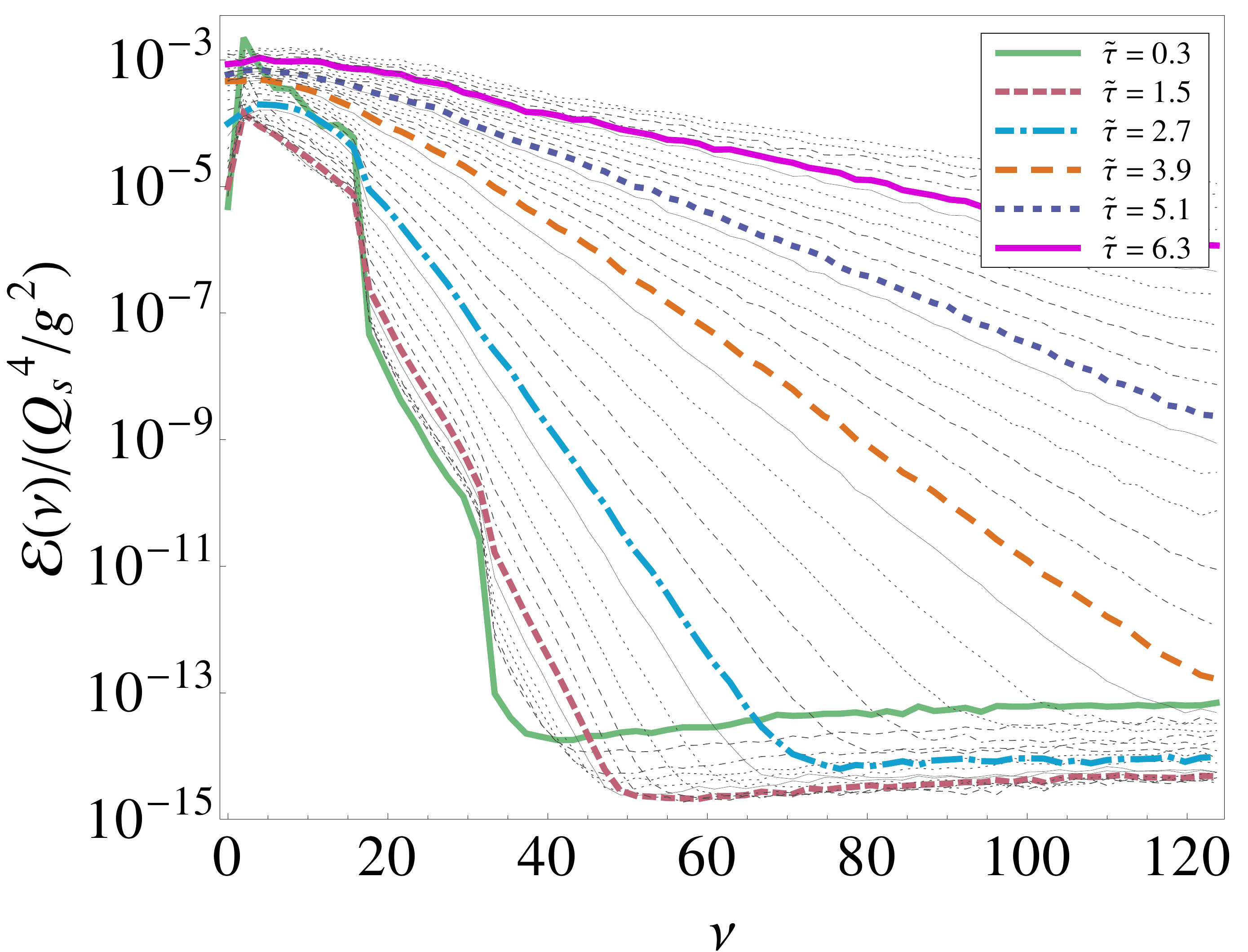}
\includegraphics[width=0.5\textwidth]{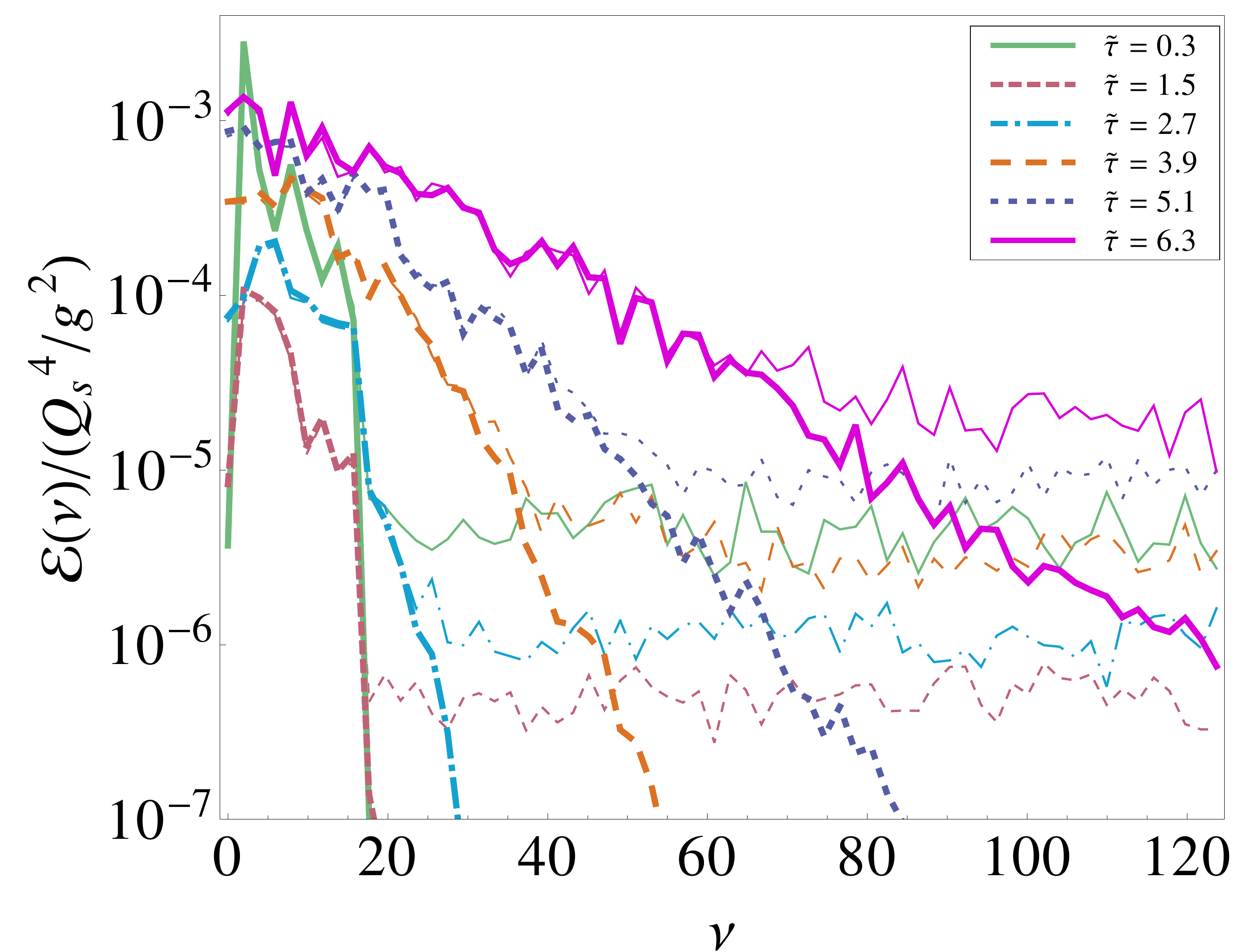}
\caption{%(Color online)
The longitudinal energy spectra at various proper times as a function of $\nu$
for the unperturbed original data (a) and the effect of a random gauge transformation on the energy spectra of a single run (b). In the latter figure
thick (thin) lines give the spectra before (after) the random gauge transformation which is as strong as possible without introducing numerical errors
in the total (gauge invariant) energy beyond 2\%.
}
\label{fig:spectra1}
\end{figure}
%%%%%%%%%%%%%%%%%%%%%%%%%%%%%%%%%%%%%%%%%%%%%%%%%%%%%%%%

The left plot of Fig.~\ref{fig:spectra1} shows our extracted
run-averaged longitudinal energy spectra at different proper times as a
function of the longitudinal wave number $\nu$ for the original data. 
The right plot of Fig.~\ref{fig:spectra1} shows one of the individual
runs before and after being subjected to random gauge transformations
as described.
We find that we can indeed produce a certain amount of noise
in the UV part of the spectrum, while the shape of the
spectrum at smaller wave numbers remains intact up to the point
where it gets drowned in the UV noise.
We take this as an indication that the energy spectra extracted 
according to (\ref{eq:nusum}) and thereby following Ref.\
\cite{Fukushima:2011nq} are reasonably robust for drawing physical
conclusions. 

However, we also found that those spectra are
not completely safe---by making the amplitude of the random
gauge transformations fully unconstrained---we could eventually
distort the spectrum also at smaller wave numbers, but
this is then accompanied by substantial violations of the
sum rule provided by the gauge invariant total energy.
We in fact intend to investigate gauge issues in more detail in the future
by also studying the effect of gauge fixing in appropriately
modified (comoving)
Coulomb gauge. %The latter should in fact work towards minimizing
%UV noise in gauge modes and thus should allow us to check that
%our spectra in temporal gauge do 

%....

\begin{figure}[t]
\hspace{4cm}
{\footnotesize (a)}
\hspace{7cm}
{\footnotesize (b)}\\
\includegraphics[width=0.45\textwidth]{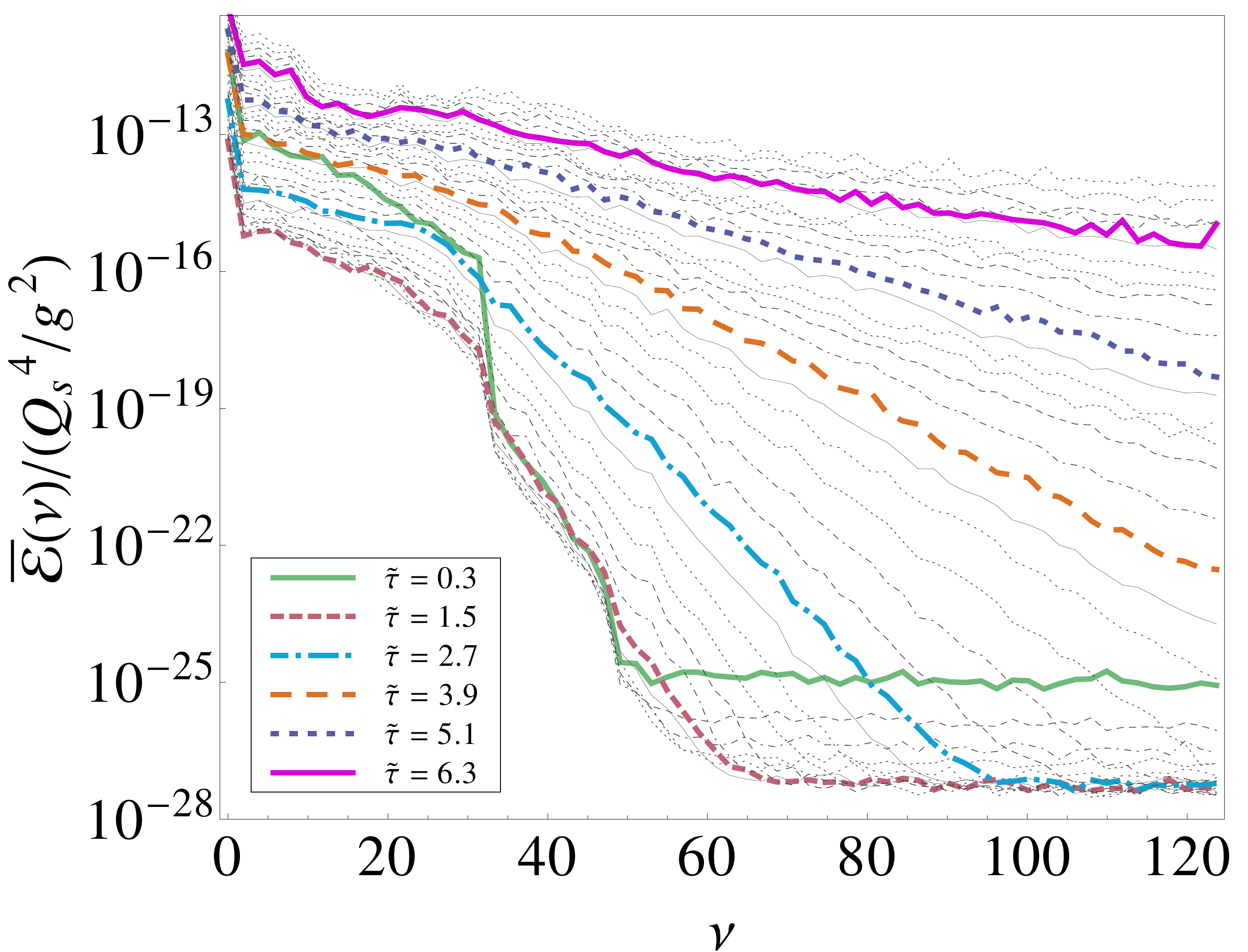}
\includegraphics[width=0.5\textwidth]{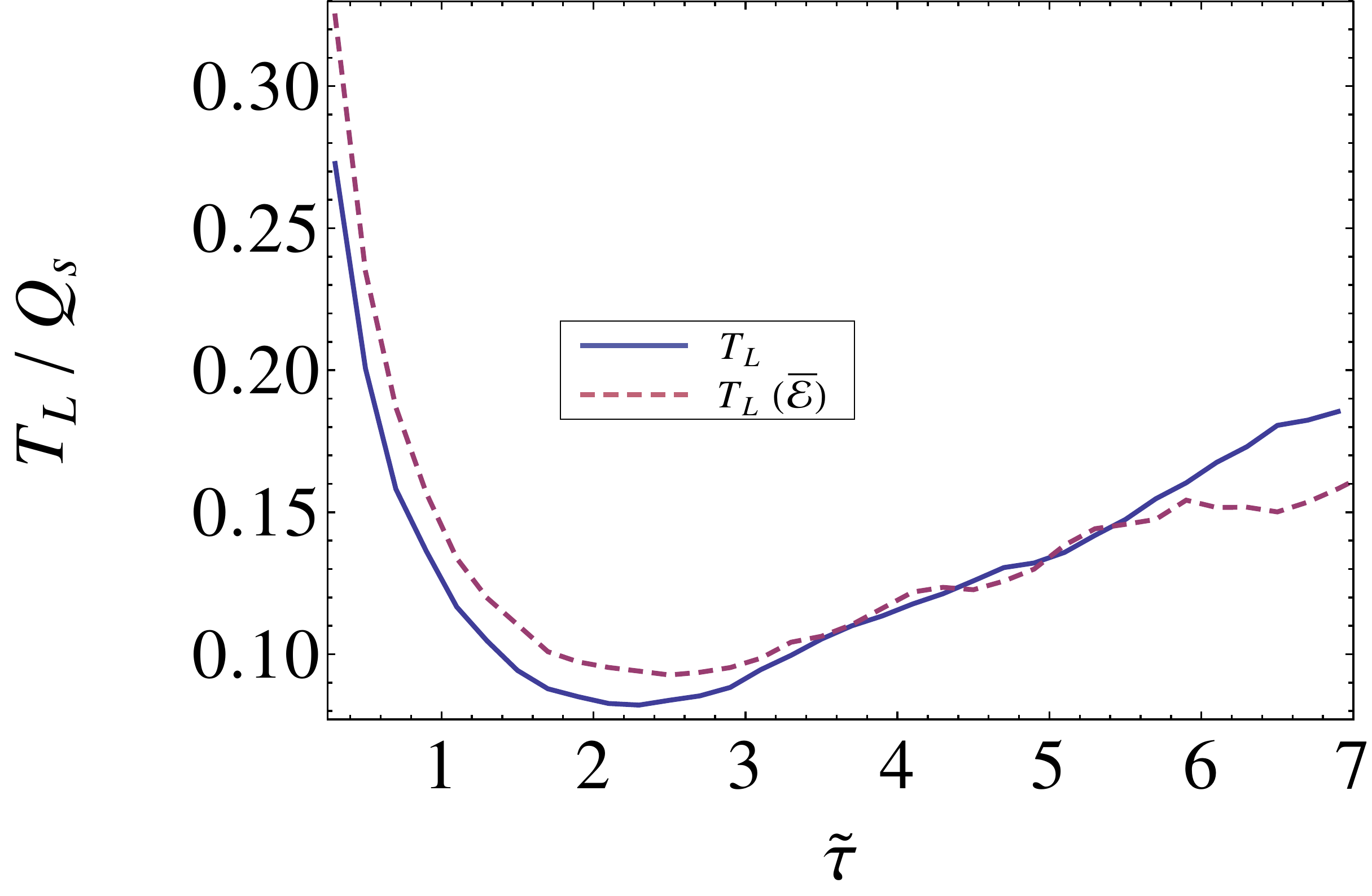}
\caption{%(Color online)
(a) The spectrum of spatial fluctuations $\bar{\mathcal E}$
in the gauge-invariant total energy
as function of longitudinal wave number $\nu$, which besides a dominant
zero-mode also shows exponential behavior; (b) The time dependent longitudinal temperature extracted from the energy spectra $\mathcal E(\nu)$
from field strength modes as
shown in Fig.~\protect\ref{fig:spectra1}a (blue full line) and compared to
that from spatial fluctuations $\bar{\mathcal E}$ (dashed red line).}
\label{fig:spectra3}
\end{figure}

As an independent check we have also analyzed the spectrum
of fluctuations in the spatial distribution of the manifestly
gauge invariant total energy. Besides an overwhelming zero
mode, this also shows an exponential behavior in longitudinal
wave number, displayed in Fig.\ \ref{fig:spectra3}a.

%The energy spectra that we have obtained (in temporal gauge)
%drop exponentially with wave number $\nu$.
One can extract a longitudinal temperature by fitting the
longitudinal energy density at each time step with a massless Boltzmann
distribution which has been integrated over transverse momenta
\begin{align}
{\cal E} \propto \int d k_z d^2 k_T \, \sqrt{k_T^2 + k_z^2} \, \exp\left(-\sqrt{k_T^2 + k_z^2}/T\right) 
\nonumber \propto \int d k_z \left( k_z^2 + 2 |k_z| T + 2 T^2 \right) \exp\left(-|k_z|/T\right) \, .
\end{align}
%To do so we used the fit function
%${\mathcal E}_{\rm fit}(k_z) = A \left( k_z^2 + 2 |k_z| T + 2 T^2 \right) \exp\left(-|k_z|/T\right),
%$
%with $A$ and $T$ as fit parameters. The result is a longitudinal
%temperature which evolves in time as shown by the full line in Fig.\ \ref{fig:spectra3}a.

In Fig.\ \ref{fig:spectra3}b we show the evolution of the
longitudinal temperature
extracted from the decomposition of the energy according to
the wave number $\nu$ in chromo-electromagnetic field strength
%according to (\ref{eq:nusum}) 
(solid line) and compare with the one obtained from the
spatial fluctuations in the gauge-invariant total energy (dashed line).
The two are seen to be remarkably similar.

%%%%%%%%%%%%%%%%%%%%%%%%%%%%%%%%%%%%%%%%%%%%%%%%%%%%%%%%%
%%%%%%%%%%%%%%%%%%%%%%%%%%%%%%%%%%%%%%%%%%%%%%%%%%%%%%%%

\section{Conclusion}

In contrast to previous simulations of anisotropic plasmas in
the hard-loop regime we have found that in a free-streaming expansion
there appears to be no saturation of growth of chromo-Weibel instabilities
and no formation of a turbulent power-law spectrum. Instead we have
observed a distribution of energy in longitudinal wave numbers which
is well described by a Boltzmann distribution with increasing temperature
as the instabilities work towards isotropization. The latter we found
to occur not before $\tilde\tau\sim 5$ (i.e.\ $\sim 10$ fm/c
at RHIC energies and $\sim 5$ fm/c at the LHC).

In addition to these results, published in
 \cite{Attems:2012js}, we have presented new results on the
scaling of the isotropization time with initial fluctuations, and
we have investigated the gauge dependence of the longitudinal
energy spectra extracted from
Fourier modes of chromo-electromagnetic fields. 
While we have found the latter to be reasonably
robust under random gauge transformations, and also remarkably
similar to the spectrum of 
spatial fluctuations in the (manifestly gauge-invariant) total energy
density, we intend to
study the mode spectra in more detail in the future, including
their form in suitably adapted Coulomb gauge so as to minimize
ultraviolet noise in pure gauge modes.

\acknowledgments 
We would like to thank the organizers of `confx' and %the participants
in particular Peter Arnold and Aleksi Kurkela for useful discussions,
in which Peter suggested to test the robustness of our results
under random gauge transformations.
We thank the Vienna Scientific Cluster under project no.\ 70061 and the LOEWE
cluster from the Center for Scientific Computing (CSC) of the Goethe University
Frankfurt under hfftheo group for providing computational resources.
M.A acknowledges funding of the Helmholtz Young Investigator Group VH-NG-822.
M.S. was supported by NSF grant no.~PHY-1068765 and the Helmholtz International
Center for FAIR LOEWE program.

\end{document}